\documentclass[article, aps, prd, amsmath, superscriptaddress, secnumarabic, nofootinbib, 10pt]{revtex4-2}
\usepackage{bm, booktabs, graphicx, hyperref, orcidlink}
\usepackage[dvipsnames]{xcolor}
\hypersetup{colorlinks=true,citecolor=RoyalBlue,linkcolor=RoyalBlue,urlcolor=RoyalBlue}
\graphicspath{{./figures/}}

\def \IITGn     {Department of Physics, Indian Institute of Technology Gandhinagar, Gujarat 382355, India.\vspace*{4pt}}

\def \Sorbonne    {Laboratoire de Physique Theorique et Hautes \'Energies (LPTHE), UMR 7589, Sorbonne Universit\'e \& CNRS, 4 place Jussieu, 75252 Paris Cedex 05, France}

\begin{document}
\title{Flavour-Changing Neutral Current Top Decays in the Three Higgs Doublet Model}

\author{\textsc{Baradhwaj Coleppa}\orcidlink{0000-0002-8761-3138}}
\email{baradhwaj@iitgn.ac.in }
\affiliation{\IITGn}

\author{\textsc{Benjamin Fuks}\orcidlink{0000-0002-0041-0566}\vspace*{7pt}}
\email{fuks@lpthe.jussieu.fr}
\affiliation{\Sorbonne}

\author{\textsc{Akshat Khanna}\orcidlink{0000-0002-2322-5929}}
\email{khanna\textunderscore akshat@iitgn.ac.in}
\affiliation{\IITGn}

\begin{abstract}
  We study flavour-changing neutral current decays of the top quark in the democratic Three Higgs Doublet Model featuring a $Z_3$-symmetric scalar potential and Natural Flavour Conservation. In this framework, while such processes are absent at tree-level, the extended scalar sector induces new one-loop contributions to rare top decays. We compute the branching ratios for processes of the form $t \to q X$ (with $q = u, c$ and $X$ denoting a boson of the model), and explore the viable regions of the parameter space under theoretical consistency conditions and current experimental constraints. Several alignment-limit scenarios corresponding to different hierarchies among the CP-even Higgs states are analysed, and we find that the predicted branching ratios can significantly exceed their Standard Model expectations while remaining consistent with existing limits. In particular, we identify scenarios with light non-standard scalars that can lead to rates within the projected sensitivity of the High-Luminosity LHC. Our results therefore highlight rare top decays as a promising probe of the extended scalar sector of the Three Higgs Doublet Model.
\end{abstract}

\maketitle


\section{Introduction}

The Standard Model (SM) of particle physics provides a remarkably successful description of the known elementary particles and their interactions. The discovery of the Higgs boson with a mass of about 125 GeV by the ATLAS and CMS collaborations~\cite{CMS:2012qbp, ATLAS:2012yve} completed the particle content of the SM, and subsequent studies of the Higgs boson's properties have shown remarkable agreement with the SM predictions~\cite{ATLAS:2022vkf, CMS:2022dwd}. Despite this success, however, the SM is widely regarded as an effective theory as it leaves several conceptual and phenomenological questions unanswered. These considerations, together with, so far, the absence of clear signals of new particles and phenomena, motivate ongoing searches for physics beyond the Standard Model (BSM), particularly at the Large Hadron Collider (LHC). Searches for BSM physics can be broadly classified into two complementary approaches. Direct searches aim at the production of new particles associated with extended scalar, gauge or matter sectors, while indirect searches probe rare processes that are highly suppressed within the SM but can receive sizeable contributions from new physics. Among such indirect probes, flavour-changing neutral current (FCNC) processes involving the top quark are specifically appealing.

Owing to its large mass and correspondingly strong coupling to the electroweak symmetry-breaking sector, the top quark indeed occupies a special role in the SM, and is thus expected to be well connected to BSM physics~\cite{Atwood:2000tu}. Within the SM, its FCNC interactions are absent at tree level and arise only through loop-induced processes involving $W$ bosons and other fermions. These amplitudes are however suppressed by the Glashow-Iliopoulos-Maiani mechanism, resulting in extremely small predicted branching ratios for the decays of the top quark into a lighter quark and an electrically-neutral gauge or Higgs boson, typically in the range $10^{-17}-10^{-12}$ depending on the channel~\cite{Eilam:1990zc, Mele:1998ag, Aguilar-Saavedra:2002lwv, Aguilar-Saavedra:2004mfd, Zhang:2013xya, Durieux:2014xla}. As a consequence, any experimental observation of top-quark FCNC processes would constitute a clear signal of new physics, and such processes are therefore actively searched for at current experiments. At the LHC, top-quark FCNC interactions have been probed both in top-antitop production when one top quark undergoes a rare decay, and in single-top production channels. While no deviation from the SM has been observed so far, the increasing dataset collected by the ATLAS and CMS experiments has led to significantly improved upper bounds on the corresponding branching ratios. The most stringent current limits are
\begin{equation}
\begin{split}
  \mathcal{BR}(t\rightarrow c\gamma) & < 4.2 \times 10^{-5}, \qquad
  \mathcal{BR}(t\rightarrow u\gamma) < 0.85 \times 10^{-5}, \\
  \mathcal{BR}(t\rightarrow cg) & < 3.7 \times 10^{-4}, \qquad
  \mathcal{BR}(t\rightarrow ug) < 0.61 \times 10^{-4}, \\
  \mathcal{BR}(t\rightarrow cZ) & < 1.3 \times 10^{-4}, \qquad
  \mathcal{BR}(t\rightarrow uZ) < 6.2 \times 10^{-5}, \\
  \mathcal{BR}(t\rightarrow ch) & < 3.7 \times 10^{-4}, \qquad
  \mathcal{BR}(t\rightarrow uh) < 1.9 \times 10^{-4},
\end{split}
\end{equation}
as reported by the two LHC collaborations~\cite{ATLAS:2021amo, ATLAS:2022per, ATLAS:2022gzn, ATLAS:2023qzr, ATLAS:2023ujo, ATLAS:2024mih, CMS:2021hug, CMS:2023bjm, CMS:2024ubt}, and the onset of the HL-LHC and advanced machine-learning methods is expected to further improve the sensitivity to these rare decay modes~\cite{Fuks:2025qgh}. From a theoretical perspective, FCNC decays of the top quark provide a powerful probe of BSM scenarios in which additional particles contribute to loop-induced amplitudes, potentially enhancing the branching ratios by many orders of magnitude. Such effects have been extensively studied in a wide range of models, including Two Higgs Doublet models~\cite{Atwood:1996vj, Botella:2015hoa, Abbas:2015cua, Baum:2008qm, Balaji:2020qjg}, the minimal supersymmetric standard model~\cite{Dedes:2014asa, Cao:2007dk, Lopez:1997xv, Eilam:2001dh}, left-right (super)symmetric~\cite{Gaitan:2004by, Frank:2005vd, Frank:2023fkc} and extra dimensional models~\cite{Gao:2013fxa, Dey:2016cve, Diaz-Furlong:2016ril}, as well as dark matter~\cite{Agrawal:2014aoa, Agrawal:2015kje, Jueid:2024cge}, quark singlet extended~\cite{Diaz-Cruz:1989tem} and composite settings~\cite{Balaji:2021lpr, Crivellin:2022fdf}.

In this work, we investigate top-quark FCNC decays within the framework of the Three Higgs Doublet Model (3HDM)~\cite{Grossman:1994jb} that features an attractive possibility for CP violation~\cite{Logan:2020mdz, Akeroyd:2021fpf} and a rich phenomenology~\cite{Dey:2023exa, CarcamoHernandez:2022vjk, Batra:2025amk} as the scalar sector of this model contains three CP-even Higgs bosons, two CP-odd Higgs bosons and four charged Higgs bosons. Tree-level FCNCs are avoided by imposing Natural Flavour Conservation (NFC)~\cite{Glashow:1976nt, Paschos:1976ay} through a discrete $Z_3$ symmetry and a democratic structure of the Yukawa sector~\cite{Cree:2011uy, Akeroyd:2016ssd}, in which each Higgs doublet couples to a single fermion type. We analyse the phenomenologically viable parameter space of the model under theoretical consistency conditions and current experimental constraints, and compute the resulting branching ratios for rare top-quark decays. Particular attention is paid to several alignment-limit scenarios \cite{Pilaftsis:2016erj, Das:2019yad, Darvishi:2021txa} corresponding to different mass hierarchies among the CP-even Higgs states.

The rest of this paper is organised as follows. In Section~\ref{sec:model} we introduce the $Z_3$-symmetric 3HDM and define its parameter space, the alignment limit and the different CP-even Higgs mass hierarchies examined. Our calculation of top-quark FCNC decays and the corresponding numerical results are next presented in Section~\ref{sec:FCNC}. Finally, we summarise our findings and conclude in Section~\ref{sec:conclusion}.

\section{The 3HDM with a \texorpdfstring{$Z_3$}~ Symmetry and Natural Flavour Conservation} \label{sec:model}

In this section, we introduce the specific realisation of the 3HDM considered in this work. The model is defined by a $Z_3$-symmetric scalar potential and a Yukawa sector implementing Natural Flavour Conservation through a democratic assignment of the fermion couplings. This structure ensures the absence of flavour-changing neutral currents at tree-level, while allowing for a rich scalar spectrum that can induce sizeable loop-level effects in rare top-quark  decay processes. We begin by describing the scalar sector of the model in section~\ref{sec:model_scalar}, including its field content, the associated scalar potential and the definition of the resulting mass eigenstates. Moreover, we devote special attention to the alignment limit which plays a central role in identifying a SM-like Higgs boson within the extended scalar spectrum. We then discuss the Yukawa sector and the resulting Higgs-fermion interactions in section~\ref{sec:model_yukawa}.

Since a detailed discussion of the full parameter space of the 3HDM and the impact of existing constraints can be found in Ref.~\cite{Batra:2025amk}, we only summarise here the essential features needed for our analysis of top-quark flavour-changing neutral current decays.

\subsection{The Scalar Potential and Mass Spectrum}\label{sec:model_scalar}

In the 3HDM, the scalar sector of the SM is extended by introducing two additional $SU(2)_L$ doublets with hypercharge $Y=1/2$. The three scalar doublets are then denoted by $\Phi_k$ ($k=1,2,3$) and are parametrised as
\begin{equation}\renewcommand{\arraystretch}{1.5}
		\Phi_k = \begin{pmatrix}
			\phi_k^+ \\ \frac{v_k + h_k + i a_k}{\sqrt{2}}
		\end{pmatrix},
  \label{eq:phi_form}
\end{equation}
where $v_k$ are the vacuum expectation values (vevs) of the neutral component of the doublets, $h_k$ denote the CP-even neutral component fields, $a_k$ the CP-odd ones and $\phi_k^+$ the charged degrees of freedom. Electroweak symmetry breaking occurs when the neutral components of the three doublets acquire non-zero vevs that satisfy $v^2 = v_1^2 + v_2^2 + v_3^2 = (246~\text{GeV})^2$. The relative magnitudes of these three vevs are conveniently parametrised by two angles $\beta_1$ and $\beta_2$ defined by $\tan\beta_1 = v_2/v_1$ and $\tan\beta_2 = v_3/\sqrt{v_1^2 + v_2^2}$. We next impose a discrete $Z_3$ symmetry under which the scalar doublets transform as $\Phi_i\to e^{i \alpha_k} \Phi_k$, where the different phases are determined by the parameter $\omega = e^{2\pi i/3}$,
\begin{equation}\label{eq:phitrans}
  \Phi_1 \rightarrow \omega \Phi_1, \qquad
  \Phi_2 \rightarrow \omega^2 \Phi_2 \qquad\text{and}\qquad
  \Phi_3 \rightarrow \Phi_3.
\end{equation}
This symmetry constrains the scalar interactions and, as discussed below, allows for the implementation of Natural Flavour Conservation in the fermion sector such that each Higgs doublet couples to a specific fermion generation.

The most general renormalisable scalar potential invariant under the $SU(2)_L \times U(1)_Y$ gauge symmetry and this $Z_3$ symmetry can be written in a compact form as
\begin{equation}\label{eq:scalarpot}\begin{split}
  V = &\ \sum_i \Big[ m_i^2\, (\Phi_i^\dagger \Phi_i) + \lambda_i\,  (\Phi_i^\dagger \Phi_i)^2\Big]
    + \sum_{i<j} \Big[\lambda_{ij}\, (\Phi_i^\dagger \Phi_i)(\Phi_j^\dagger \Phi_j) + \lambda'_{ij}\, (\Phi_i^\dagger \Phi_j)(\Phi_j^\dagger \Phi_i) \Big]\\
    &\qquad + \Big[\tilde\lambda_1\, (\Phi_1^\dagger\Phi_2)(\Phi_1^\dagger\Phi_3) + \tilde\lambda_2\, (\Phi_1^\dagger\Phi_2)(\Phi_3^\dagger\Phi_2) + \tilde\lambda_3\, (\Phi_1^\dagger\Phi_3)(\Phi_2^\dagger\Phi_3) + \mathrm{H.c.} \Big],\quad
\end{split}\end{equation}
where $i,j\in \{1,2,3\}$, and the terms in the final line specifically represent the $Z_3$-invariant couplings between all three doublets together. In our notation, we denote by $m_i^2$ the three quadratic mass parameters, while $\lambda$, $\lambda'$ and $\tilde\lambda$ encode the strength of the various quartic couplings. In general, the $\tilde\lambda$ parameters may be complex, potentially introducing CP violation in the scalar sector. In this work, however, we assume a CP-conserving scenario by taking all potentially complex couplings and vevs to be real. As a result, the physical neutral scalar eigenstates can be classified according to their CP quantum numbers, and no mixing occurs between the CP-even and CP-odd states.

After electroweak symmetry breaking, the model contains three CP-even neutral Higgs bosons $H_{1,2,3}$, two CP-odd neutral states $A_{2,3}$ and two charged Higgs bosons $H_{2,3}^\pm$, together with the Goldstone modes $G^0$ and $G^\pm$ which are absorbed by the $Z$ and $W^\pm$ gauge bosons. The CP-even squared mass matrix, obtained from the second derivatives of the scalar potential in Eqn.~\eqref{eq:scalarpot} evaluated at the electroweak vacuum, is a real symmetric $3\times3$ matrix $\mathcal{M}^2_S$. Its independent components are given by
\begin{equation}\renewcommand{\arraystretch}{1.3}
\mathcal{M}^2_S = \scalebox{0.88}{$\begin{pmatrix} 
  2\lambda_{1}v_1^2 - \frac{v_2v_3}{2v_1} \left[ \tilde{\lambda}_{2}v_2 + \tilde{\lambda}_{3}v_3 \right] & (\lambda_{12} + \lambda'_{12})v_1v_2 + \tilde{\lambda}_{1}v_1v_3 + \tilde{\lambda}_{2}v_2v_3 + \frac{1}{2}\tilde{\lambda}_{3}v_3^2 & (\lambda_{13} + \lambda'_{13})v_1v_3 + \tilde{\lambda}_{1}v_1v_2 + \tilde{\lambda}_{3}v_2v_3 + \frac{1}{2}\tilde{\lambda}_{2}v_2^2 \\
   & 2\lambda_{2}v_2^2 - \frac{v_1v_3}{2v_2} \left[ \tilde{\lambda}_{1}v_1 + \tilde{\lambda}_{3}v_3 \right] & (\lambda_{23} + \lambda'_{23})v_2v_3 + \tilde{\lambda}_{2}v_1v_2 + \tilde{\lambda}_{3}v_1v_3 + \frac{1}{2}\tilde{\lambda}_{1}v_1^2  \\
   &  & 2\lambda_{3}v_3^2 - \frac{v_1v_2}{2v_3} \left[ \tilde{\lambda}_{1}v_1 + \tilde{\lambda}_{2}v_2 \right]
\end{pmatrix}$},\end{equation}
where the lower triangular block is omitted for brevity, and we used the minimisation conditions of the potential to eliminate the dependence on the quadratic parameters $m_i^2$. This matrix is diagonalised by an orthogonal transformation $O_\alpha$ such that $O_\alpha\, \mathcal{M}^2_S\, O_\alpha^T = \mathrm{diag}(m_{H 1}^2, m_{H 2}^2, m_{H 3}^2)$. Following the standard parametrisation for a $3\times 3$ rotation matrix, $O_\alpha$ is defined by three mixing angles $\alpha_{1,2,3}$, 
\begin{equation}
    \label{eq:matalphtrans}
    O_\alpha = \begin{pmatrix}
        c_{\alpha_1} c_{\alpha_2} & c_{\alpha_2} s_{\alpha_1} & s_{\alpha_2} \\ 
        -c_{\alpha_3} s_{\alpha_1} - s_{\alpha_3} s_{\alpha_2} c_{\alpha_1} & c_{\alpha_3} c_{\alpha_1} - s_{\alpha_3} s_{\alpha_2} s_{\alpha_1} & s_{\alpha_3} c_{\alpha_2} \\ 
        s_{\alpha_3} s_{\alpha_1} - c_{\alpha_3} s_{\alpha_2} c_{\alpha_1} & -s_{\alpha_3} c_{\alpha_1} - c_{\alpha_3} s_{\alpha_2} s_{\alpha_1} & c_{\alpha_3} c_{\alpha_2} 
    \end{pmatrix},
\end{equation}
where $c_{\alpha_i}$ and $s_{\alpha_i}$ refer to the cosine and sine of the angle $\alpha_i$ respectively. The physical mass eigenstates $H_{1,2,3}$ are then related to the gauge eigenstates $h_{1,2,3}$ through the relation $H_i = (O_\alpha)_{ij} h_j$. While the physical scalar masses $m_{H_{1,2,3}}$ are formally functions of the quartic couplings, in this work we treat them and the mixing angles $\alpha_{1,2,3}$ as independent input parameters, together with the mass and mixing angles associated with the rest of the Higgs sector and introduced below. The underlying potential parameters are hence numerically determined in terms of these physical observables.

The charged scalar sector, which contains after electroweak symmetry breaking two pairs of physical charged Higgs bosons $H_{2,3}^\pm$ and the Goldstone modes $G^\pm$, is treated similarly. The $3\times 3$ real symmetric squared mass matrix $\mathcal{M}^2_\pm$ associated with these degrees of freedom is extracted from the scalar potential in Eqn.~\eqref{eq:scalarpot}, and reads
\begin{equation}\renewcommand{\arraystretch}{1.8}
  \mathcal{M}^2_\pm = \scalebox{0.805}{$\begin{pmatrix} 
  -\frac{v_2^2}{2}\lambda'_{12} \!-\! \frac{v_3^2}{2}\lambda'_{13} \!-\! \tilde{\lambda}_{1}v_2v_3 \!-\! \frac{v_2v_3}{2v_1}(\tilde{\lambda}_{2}v_2\!+\!\tilde{\lambda}_{3}v_3) & \frac{1}{2}\left( v_1v_2\lambda'_{12} + v_1v_3\tilde{\lambda}_{1} + v_2v_3\tilde{\lambda}_{2} \right) &  \frac{1}{2}\left( v_1v_2\tilde{\lambda}_{1} + v_1v_3\lambda'_{13} + v_2v_3\tilde{\lambda}_{3} \right)\\
  & -\frac{v_1^2}{2}\lambda'_{12} \!-\! \frac{v_3^2}{2}\lambda'_{23} \!-\! \tilde{\lambda}_{2}v_1v_3 \!-\! \frac{v_1v_3}{2v_2}(\tilde{\lambda}_{1}v_1\!+\!\tilde{\lambda}_{3}v_3) & \frac{1}{2}\left( v_1v_2\tilde{\lambda}_{2} + v_1v_3\tilde{\lambda}_{3} + v_2v_3\lambda'_{23} \right) \\
  &  & -\frac{v_1^2}{2}\lambda'_{13} \!-\! \frac{v_2^2}{2}\lambda'_{23} \!-\! \tilde{\lambda}_{3}v_1v_2 \!-\! \frac{v_1v_2}{2v_3}(\tilde{\lambda}_{1}v_1\!+\!\tilde{\lambda}_{2}v_2)
\end{pmatrix}$},\end{equation}
where the lower triangular block can once again be obtained by symmetry, and where we used the minimisation conditions of the potential to eliminate the dependence on the quadratic parameters $m_i^2$. As required by gauge invariance, this matrix possesses one zero eigenvalue corresponding to the charged Goldstone mode. Its diagonalisation is then performed in two steps. First, we rotate the gauge eigenstates into the Higgs basis using the matrix $O_\beta$, which depends on the vev angles $\beta_{1,2}$ and is given by
\begin{equation}\renewcommand{\arraystretch}{1.3}
    O_\beta = \begin{pmatrix}
        c_{\beta_2} c_{\beta_1} & c_{\beta_2} s_{\beta_1} & s_{\beta_2} \\ 
        -s_{\beta_1} & c_{\beta_1} & 0 \\ 
        -c_{\beta_1} s_{\beta_2} & -s_{\beta_1} s_{\beta_2} & c_{\beta_2}
    \end{pmatrix}.
\end{equation}
This similarity transformation isolates the massless Goldstone boson $G^\pm$ in the first component of the rotated field vectors. The remaining $2 \times 2$ block is diagonalised by a subsequent rotation $O_\gamma$ depending on a mixing angle $\gamma$,
\begin{equation}
    O_\gamma = \begin{pmatrix} 1 & 0 & 0 \\ 0 & c_\gamma & -s_\gamma \\ 0 & s_\gamma & c_\gamma \end{pmatrix}.
\end{equation}
The physical mass eigenstates are then given by the total transformation $O_C = O_\gamma O_\beta$, such that $\text{diag}(0, m_{H^\pm_2}^2, m_{H^\pm_3}^2) = O_C \mathcal{M}^2_{\phi^{\pm}} O_C^T$. Explicitly, the relation between mass and gauge eigenstates is given by
\begin{equation}
    \label{chargetransrelat}
    \begin{split}
        G^\pm & = c_{\beta_1}c_{\beta_2} \phi_1^\pm + c_{\beta_2}s_{\beta_1} \phi_2^\pm + s_{\beta_2} \phi_3^\pm, \\
        H_2^\pm & = (-c_{\gamma }s_{\beta_1} + c_{\beta_1}s_{\beta_2}s_{\gamma})\phi_1^\pm + (c_{\beta_1}c_{\gamma} + s_{\beta_1}s_{\beta_2}s_{\gamma})\phi_2^\pm + (-c_{\beta_2}s_{\gamma}) \phi_3^\pm,\\
        H_3^\pm & = (-c_{\beta_1}c_{\gamma}s_{\beta_2}-s_{\beta_1}s_{\gamma})\phi_1^\pm + (-c_{\gamma}s_{\beta_1}s_{\beta_2}+c_{\beta_1}s_{\gamma})\phi_2^\pm + (c_{\beta_2}c_{\gamma}) \phi_3^\pm,
    \end{split}
\end{equation}
and as mentioned above, in our numerical analysis we treat the physical masses $m_{H^\pm_{2,3}}$ and the mixing angle $\gamma$ as independent input parameters.

Finally, we follow the same steps for the CP-odd neutral sector, which comprises two physical pseudoscalar states $A_{1,2}$ and the neutral Goldstone mode $G^0$, and extract the corresponding $3 \times 3$ real symmetric squared mass matrix $\mathcal{M}^2_A$. This matrix is given, after applying the scalar potential minimisation conditions to remove the dependence on the quadratic parameters $m^2_i$, by
\begin{equation}\renewcommand{\arraystretch}{1.8}
  \mathcal{M}^2_A = \begin{pmatrix} 
  - \frac{v_2v_3}{2v_1}(\tilde{\lambda}_{2}v_2+\tilde{\lambda}_{3}v_3) - 2 \tilde{\lambda}_1 v_2 v_3 & v_1 v_3 \tilde{\lambda}_1 + v_2 v_3 \tilde{\lambda}_2 - \frac12 \tilde{\lambda}_3 v_3^2 & v_1 v_2 \tilde{\lambda}_1 + v_2 v_3 \tilde{\lambda}_3 - \frac12 \tilde{\lambda}_2 v_2^2\\
  &- \frac{v_1v_3}{2v_2}(\tilde{\lambda}_{1}v_1+\tilde{\lambda}_{3}v_3) - 2 \tilde{\lambda}_2 v_1 v_3 & v_1 v_2 \tilde{\lambda}_2 + v_1 v_3 \tilde{\lambda}_3 - \frac12 \tilde{\lambda}_1 v_1^2\\
  && - \frac{v_1v_2}{2v_3}(\tilde{\lambda}_{1}v_1+\tilde{\lambda}_{2}v_2) - 2 \tilde{\lambda}_3 v_1 v_2
\end{pmatrix}.\end{equation}
Analogous to the charged scalar sector, the matrix $\mathcal{M}^2_{A}$ is diagonalised in two steps. We first rotate the field basis into the Higgs basis using the rotation matrix $O_\beta$ to isolate the massless Goldstone boson $G^0$, and then apply a second rotation of matrix $O_\theta$ depending on a mixing angle $\theta$ allowing for the  diagonalisation of the remaining physical $2 \times 2$ subspace. The total transformation matrix is thus given by $O_A = O_\theta O_\beta$, where
\begin{equation}
    O_\theta = \begin{pmatrix} 1 & 0 & 0 \\ 0 & c_\theta & -s_\theta \\ 0 & s_\theta & c_\theta \end{pmatrix},
\end{equation}
and it allows for the derivation of the relation between the physical mass eigenstates and the gauge eigenstates,
\begin{equation}
    \begin{split}
        G_0 & = (c_{\beta_1}c_{\beta_2}) a_1 + (c_{\beta_2}s_{\beta_1}) a_2 + (s_{\beta_2}) a_3, \\
        A_2 & = (-c_{\theta }s_{\beta_1} + c_{\beta_1}s_{\beta_2}s_{\theta})a_1 + (c_{\beta_1}c_{\theta} + s_{\beta_1}s_{\beta_2}s_{\theta})a_2 + (-c_{\beta_2}s_{\theta})a_3, \\
        A_3 & = (-c_{\beta_1}c_{\theta}s_{\beta_2}-s_{\beta_1}s_{\theta})a_1 + (-c_{\theta}s_{\beta_1}s_{\beta_2}+c_{\beta_1}s_{\theta})a_2 + (c_{\beta_2}c_{\theta}) a_3.
    \end{split}
\end{equation}
Similarly to the CP-even and charged scalars, in our study the pseudoscalar masses $m_{A_{1,2}}$ and the mixing angle $\theta$ are treated as independent input parameters, which in turn determine the values of the $\tilde{\lambda}_i$ couplings.

The observation of a Higgs boson with a mass of approximately 125~GeV and couplings in close agreement with the SM predictions strongly constrains extended Higgs sectors. In multi-Higgs-doublet models, these observations imply, most notably, that one of the CP-even neutral scalar mass eigenstates must possess SM-like couplings to the electroweak gauge bosons. This requirement defines the so-called alignment limit~\cite{Gunion:2002zf, Bernon:2015qea, Grzadkowski:2018ohf}, in which a CP-even Higgs state aligns with the direction of the electroweak vev in field space. In the 3HDM, this alignment can be realised in several ways~\cite{Pilaftsis:2016erj, Das:2019yad, Darvishi:2021txa}. Unlike in simpler BSM extensions with only two Higgs doublets, the 3HDM does not impose a predetermined mass ordering among the CP-even scalars. As a result, any of the three CP-even states may, in principle, play the role of the SM-like Higgs boson, depending on the values of the scalar potential parameters and the vacuum structure. This feature allows for three qualitatively distinct alignment scenarios, corresponding to different mass hierarchies among the CP-even states.

In scenarios defined by a regular hierarchy, the lightest CP-even scalar $H_1$ is identified with the 125~GeV SM-like Higgs boson. Alignment then requires the $H_1VV$ ($V=W,Z$) couplings to reproduce their SM values, leading to the condition
\begin{equation}
    c_{\beta_2} c_{\alpha_2} c_{\alpha_1 - \beta_1} + s_{\beta_2} s_{\alpha_2} = 1.
\end{equation}
This condition is satisfied, in particular, when $c_{\alpha_1-\beta_1} = 1$ (implying $\alpha_1 = \beta_1 + 2n\pi$) and $\alpha_2 = \beta_2+2n\pi$. While other mathematical solutions with $c_{\alpha_1-\beta_1} \neq 1$ exist, they typically correspond to redundant configurations related by field redefinitions ($c_{\alpha_1-\beta_1}=-1$), or to specific vacuum structures where electroweak symmetry breaking is driven by a single doublet ($c_{\alpha_1-\beta_1}\neq \pm 1$). Following Ref.~\cite{Batra:2025amk}, we thus restrict our analysis to the regularly aligned case with $c_{\alpha_1-\beta_1}= 1$. In a setup with a medial hierarchy, the second-lightest CP-even scalar $H_2$ is SM-like, thus situated between one lighter and one heavier CP-even states. The alignment condition for $H_2$ yields
\begin{equation}
    \pm c_{\beta_2} c_{\alpha_3} s_{\alpha_1-\beta_1}
    + s_{\beta_2} c_{\alpha_2} s_{\alpha_3}
    - c_{\beta_2} s_{\alpha_2} s_{\alpha_3} c_{\alpha_1-\beta_1} = 1.
\end{equation}
Again, focusing on the physically relevant branch where $c_{\alpha_1-\beta_1} = 1$, the condition reduces to $s_{\alpha_3} s_{\beta_2 - \alpha_2} = 1$, which can be satisfied by imposing $\alpha_1 = \beta_1 + 2n\pi$ together with $\alpha_3 = \pi/2$ or $3\pi/2$ and $\tan\alpha_2 = -\cot\beta_2$. Finally, in scenarios featuring an inverted hierarchy, the heaviest CP-even scalar $H_3$ is identified as the SM-like Higgs boson, allowing for two lighter scalars with masses below 125~GeV. The corresponding alignment condition reads
\begin{equation}
    c_{\alpha_3} s_{\beta_2} c_{\alpha_2}
    - c_{\alpha_3} c_{\beta_2} s_{\alpha_2} c_{\alpha_1-\beta_1}
    \pm c_{\beta_2} s_{\alpha_3} s_{\alpha_1-\beta_1} = 1.
\end{equation}
As in the previous cases, this is satisfied for $c_{\alpha_1-\beta_1} = 1$ and therefore $\alpha_1 = \beta_1 + 2n\pi$, which simplifies the above requirement to $c_{\alpha_3} s_{\beta_2 - \alpha_2} = 1$. This last condition can be realised by setting $\alpha_3 = 0$ or $\pi$ and imposing $\tan\alpha_2 = -\cot\beta_2$.

\begin{table}\renewcommand{\arraystretch}{1.3}{}
    \centering
    \resizebox{.99\textwidth}{!}{
    \begin{tabular}{l l l l l}
    \toprule[1pt]
        Coupling & \ \ \ \ \  Coupling & \ \ \ \ \ Regular & \ \ \ \ \ Medial & \ \ \ \ \ Inverted \\
        \midrule[1pt]
        $H_1H_2^-W^+$ & \ \ \ \ \  $c_{\beta_2}s_{\alpha_2}s_{\theta}-c_{\alpha_2}(c_{\alpha_1-\beta_1}s_{\beta_2}s_{\theta}+s_{\alpha_1-\beta_1}c_{\theta})$ & \ \ \ \ \ 0 & \ \ \ \ \ $-s_{\theta}$ & \ \ \ \ \  $-s_{\theta}$\\
        $H_2H_2^-W^+$ & \ \ \ \ \  $s_{\alpha_3}((c_{\alpha_2}c_{\beta_2}+c_{\alpha_1-\beta_1}s_{\alpha_2}s_{\beta_2})s_{\theta}+s_{\alpha_1-\beta_1}s_{\alpha_2}c_{\theta})-c_{\alpha_3}(c_{\alpha_1-\beta_1}c_{\theta}-s_{\alpha_1-\beta_1}s_{\beta_2}s_{\theta})$ & \ \ \ \ \ $-c_{(\alpha_3+\theta)}$ & \ \ \ \ \ 0 & \ \ \ \ \ $-c_{\theta}$ \\
        $H_3H_2^-W^+$ & \ \ \ \ \  $s_{\alpha_3}(-s_{\alpha_1-\beta_1}s_{\beta_2}s_{\theta}+c_{\alpha_1-\beta_1}c_{\theta})+c_{\alpha_3}(s_{\theta}(c_{\alpha_2}c_{\beta_2}+c_{\alpha_1-\beta_1}s_{\alpha_2}s_{\beta_2})+s_{\alpha_1-\beta_1}s_{\alpha_2}c_{\theta})$ & \ \ \ \ \ $s_{(\alpha_3+\theta)}$ & \ \ \ \ \ $c_{\theta}$ & \ \ \ \ \ 0 \\
        $H_1H_3^-W^+$ & \ \ \ \ \ $s_{\alpha_2}c_{\beta_2}c_{\theta}-c_{\alpha_2}(c_{\alpha_1-\beta_1}s_{\beta_2}c_{\theta}-s_{\alpha_1-\beta_1}s_{\theta})$ & \ \ \ \ \ 0 & \ \ \ \ \ $-c_{\theta}$ & \ \ \ \ \ $-c_{\theta}$ \\
        $H_2H_3^-W^+$ & \ \ \ \ \  $-s_{\alpha_3}(c_{\theta}(c_{\alpha_2}c_{\beta_2}+c_{\alpha_1-\beta_1}s_{\alpha_2}s_{\beta_2})-s_{\alpha_1-\beta_1}s_{\alpha_2}s_{\theta})+c_{\alpha_3}(-s_{\alpha_1-\beta_1}s_{\beta_2}c_{\theta}-c_{\alpha_1-\beta_1}s_{\theta})$ & \ \ \ \ \ $-s_{(\alpha_3+\theta)}$ & \ \ \ \ \ 0 & \ \ \ \ \ $-s_{\theta}$ \\
        $H_3H_3^-W^+$ & \ \ \ \ \ $s_{\alpha_3}(c_{\alpha_1-\beta_1}s_{\theta}+s_{\alpha_1-\beta_1}s_{\beta_2}c_{\theta})-c_{\alpha_3}(c_{\theta}(c_{\alpha_2}c_{\beta_2}+c_{\alpha_1-\beta_1}s_{\alpha_2}s_{\beta_2})-s_{\alpha_1-\beta_1}s_{\alpha_2}s_{\theta})$ & \ \ \ \ \ $-c_{(\alpha_3+\theta)}$ & \ \ \ \ \ $s_{\theta}$ & \ \ \ \ \ 0 \\
        $A_2H_2^-W^+$ & \ \ \ \ \ $c_{(\gamma-\theta)}$ & \ \ \ \ \ $c_{(\gamma-\theta)}$ & \ \ \ \ \ $c_{(\gamma-\theta)}$ & \ \ \ \ \ $c_{(\gamma-\theta)}$ \\
        $A_3H_2^-W^+$ & \ \ \ \ \  $s_{(\gamma-\theta)}$ & \ \ \ \ \ $s_{(\gamma-\theta)}$ & \ \ \ \ \ $s_{(\gamma-\theta)}$ & \ \ \ \ \ $s_{(\gamma-\theta)}$ \\
        $A_2H_3^-W^+$ & \ \ \ \ \  $-s_{(\gamma-\theta)}$ & \ \ \ \ \ $-s_{(\gamma-\theta)}$ & \ \ \ \ \ $-s_{(\gamma-\theta)}$ & \ \ \ \ \ $-s_{(\gamma-\theta)}$ \\
        $A_3H_3^-W^+$ & \ \ \ \ \  $c_{(\gamma-\theta)}$ & \ \ \ \ \ $c_{(\gamma-\theta)}$ & \ \ \ \ \ $c_{(\gamma-\theta)}$ & \ \ \ \ \ $c_{(\gamma-\theta)}$ \\
    \bottomrule[1pt]
    \end{tabular}}
    \caption{Trilinear gauge-Higgs-Higgs couplings in the democratic 3HDM relevant for the present analysis. The third, fourth and fifth columns examine these  couplings in scenarios featuring a regular, medial and inverted alignment limit, respectively. \label{tab:VHH_couplings}}
\end{table}

The interactions between the scalar states and the electroweak gauge bosons are strongly constrained by the structure of the model and by the alignment limit. As a consequence, the trilinear couplings between the SM-like Higgs state, a charged Higgs boson and a $W$ boson vanish identically. In contrast, the remaining two CP-even neutral scalars exhibit non-vanishing and complementary couplings to an associated pair of particles comprising a charged Higgs and a $W$ boson, as shown in Table~\ref{tab:VHH_couplings}. The precise pattern of these couplings then depends on the regular, medial or inverted hierarchy satisfied by the CP-even scalar masses, and this plays an important role in loop-induced processes as it determines which scalar states can contribute significantly once alignment is imposed.

The CP-odd scalars, on the other hand, retain unsuppressed couplings to charged Higgs bosons and $W$ bosons, that depend only on the mixing in the CP-odd sector and are insensitive to the CP-even alignment choice. As a result, both non-SM CP-even and CP-odd scalars can provide sizeable contributions to loop-induced flavour-changing neutral current processes involving the top quark.

\subsection{Yukawa Interactions and Natural Flavor Conservation}\label{sec:model_yukawa}

Flavour-Changing Neutral Currents are strongly constrained by experimental data, particularly in the quark sector. On the other hand, multi-Higgs-doublet models generically predict tree-level FCNCs because multiple scalar doublets can couple to the same fermion type, leading to non-diagonal Yukawa matrices that cannot be simultaneously diagonalised with the mass matrices. To suppress these dangerous transitions, we invoke the principle of Natural Flavour Conservation~\cite{Glashow:1976nt, Paschos:1976ay}, which, according to the Paschos-Glashow-Weinberg theorem, is satisfied if all fermions of a given charge couple to exactly one specific scalar doublet. In the 3HDM scenarios explored in this work, NFC is achieved by extending the discrete $Z_3$ symmetry of Eqn.~\eqref{eq:phitrans} to the fermion sector, adopting a Type-Z or democratic Yukawa structure where each fermion class couples to a distinct doublet~\cite{Cree:2011uy, Akeroyd:2016ssd}. Specifically, we assign the $Z_3$ charges such that the up-type quark singlets couple to $\Phi_3$, the down-type quark singlets to $\Phi_2$ and the charged lepton singlets to $\Phi_1$. This ensures that all Yukawa matrices are proportional to the fermion mass matrices, thereby eliminating FCNCs at tree level. Under the $Z_3$ symmetry, the right-handed fermions thus transform as
\begin{equation} \label{eq:fermtrans}
  d_R  \rightarrow \omega d_R , \qquad 
  \ell_R \rightarrow \omega^2 \ell_R , \qquad 
  u_R \rightarrow u_R,
\end{equation}
whereas the left-handed doublets $Q_L$ and $L_L$ transform trivially. The resulting Yukawa Lagrangian reads
\begin{equation}
		\label{eq:yukeq}
		\mathcal{L}_\mathrm{Yukawa} = - \bigg[\bar{L}_L \Phi_1 \mathcal{Y}_\ell \ell_R+\bar{Q}_L \Phi_2 \mathcal{Y}_d d_R + \bar{Q}_L \tilde{\Phi}_3 \mathcal{Y}_u u_R  + \mathrm{H.c.} \bigg],
\end{equation}
where $\mathcal{Y}_{\ell, d, u}$ represent the Yukawa matrices and $\tilde{\Phi}_3 = i \sigma_2 \Phi_3^*$. Consequently, the Yukawa matrices can be expressed in terms of the fermion mass matrices $\mathcal{M}_f$ as 
\begin{equation}
  \mathcal{Y}_\ell = \frac{\sqrt{2}\, \mathcal{M}_\ell}{v_1},\qquad
  \mathcal{Y}_d = \frac{\sqrt{2}\, \mathcal{M}_d}{v_2},\qquad
  \mathcal{Y}_u = \frac{\sqrt{2}\, \mathcal{M}_u}{v_3}.
\end{equation}

\begin{table}\renewcommand{\arraystretch}{1.3}{}
    \centering
    \begin{tabular}{l l l l l}
    \toprule[1pt]
        Coupling & \ \ \ \ \  Scaling factor & \ \ \ \ \ Regular & \ \ \ \ \ Medial & \ \ \ \ \ Inverted \\
        \midrule[1pt]
        ${\bar{b} b} {H_1}$ & \ \ \ \ \  $\frac{s_{\alpha_1}c_{\alpha_2}}{s_{\beta_1}c_{\beta_2}}$ &  \ \ \ \ \ 1 & \ \ \ \ \ $t_{\beta_2}$ &  \ \ \ \ \ $t_{\beta_2}$ \\
        ${\bar{b} b} {H_2}$ & \ \ \ \ \  $-\frac{(c_{\alpha_1}c_{\alpha_3}-s_{\alpha_1}s_{\alpha_2}s_{\alpha_3})}{s_{\beta_1}c_{\beta_2}}$ & \ \ \ \ \ $\frac{(c_{\beta_1}c_{\alpha_3}-s_{\beta_1}s_{\beta_2}s_{\alpha_3})}{s_{\beta_1}c_{\beta_2}}$ &  \ \ \ \ \ 1 & \ \ \ \ \ $-\frac{c_{\beta_1}}{s_{\beta_1}c_{\beta_2}}$ \\
        ${\bar{b} b} {H_3}$ & \ \ \ \ \  $\frac{(c_{\alpha_1}s_{\alpha_3}+s_{\alpha_1}s_{\alpha_2}c_{\alpha_3})}{s_{\beta_1}c_{\beta_2}}$ & \ \ \ \ \ $\frac{(c_{\beta_1}s_{\alpha_3}+s_{\beta_1}s_{\beta_2}c_{\alpha_3})}{s_{\beta_1}c_{\beta_2}}$ & \ \ \ \ \ $\frac{c_{\beta_1}}{s_{\beta_1}c_{\beta_2}}$ & \ \ \ \ \ 1 \\
        ${\bar{t} t} {H_1}$ & \ \ \ \ \  $\frac{s_{\alpha_2}}{s_{\beta_2}}$ & \ \ \ \ \ 1 & \ \ \ \ \ $-\frac{1}{t_{\beta_2}}$ & \ \ \ \ \ $-\frac{1}{t_{\beta_2}}$ \\
        ${\bar{t} t} {H_2}$ & \ \ \ \ \  $\frac{c_{\alpha_2}s_{\alpha_3}}{s_{\beta_2}}$ & \ \ \ \ \ $\frac{c_{\beta_2}s_{\alpha_3}}{s_{\beta_2}}$ & \ \ \ \ \ 1 & \ \ \ \ \ 0 \\
        ${\bar{t} t} {H_3}$ & \ \ \ \ \  $\frac{c_{\alpha_2}c_{\alpha_3}}{s_{\beta_2}}$ & \ \ \ \ \ $\frac{c_{\beta_2}c_{\alpha_3}}{s_{\beta_2}}$ & \ \ \ \ \ 0 & \ \ \ \ \ 1 \\
    \bottomrule[1pt]
    \end{tabular}
    \caption{Neutral scalar couplings to quarks in the democratic 3HDM, normalised to the SM Higgs couplings. The third, fourth and fifth columns represent these scaling factorin scenario featuring a regular, medial and inverted alignment limit, respectively. \label{tab:HFF_couplings}}
\end{table}
The couplings of the physical neutral scalars to the fermions are modified relative to the SM predictions. In the alignment limit, one CP-even neutral scalar aligns with the vev direction, recovering SM-like coupling strengths, while the remaining two CP-even states acquire complementary, non-SM couplings, exactly as for their couplings to a charged Higgs state and $W$ boson. Furthermore, the CP-odd scalars $A_{1,2}$ also couple to fermions, with strengths determined by the pseudoscalar mixing angles. Table~\ref{tab:HFF_couplings} collects expressions for the normalised Yukawa couplings (\textit{i.e.}\ the scaling factors multiplying the corresponding SM interaction) for up-type and down-type quarks across the three alignment hierarchies. The pattern demonstrates that the SM-like Higgs retains the interaction strength predicted in the SM, while the non-SM scalars carry the remaining interactions in a complementary way, a consequence of the unitarity of the scalar mixing matrices. This is crucial for loop-induced processes, as this indeed dictates which scalars dominate the amplitudes once alignment is imposed.

These Yukawa structures have direct consequences for FCNC processes, particularly those involving the top quark. Since NFC prevents tree-level FCNCs, any top-up or top-charm transitions mediated by neutral scalars can only occur at the loop level, and these amplitudes are controlled by the non-SM CP-even and CP-odd scalars. The complementarity observed in the Yukawa sector ensures that, while the SM-like Higgs retains its full SM-strength couplings, the non-SM scalars inherit the remaining interaction strength. This allows them to generate potentially sizeable loop-induced effects. Consequently, the mass hierarchy and the specific fermionic coupling structure are the primary drivers of the top FCNC magnitude in our 3HDM framework.

\section{Top Quark FCNCs in the Democratic 3HDM}
\label{sec:FCNC}

\subsection{Calculation of Top FCNCs and Technical Framework}\label{sec:framework}

\begin{figure}
   \centering
    \includegraphics[scale=0.25]{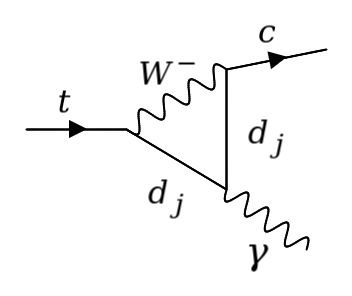}
    \includegraphics[scale=0.25]{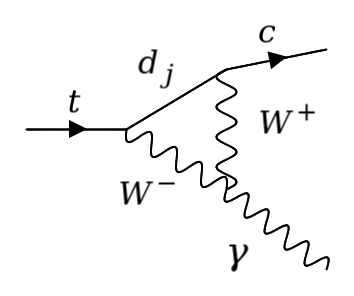}
    \includegraphics[scale=0.25]{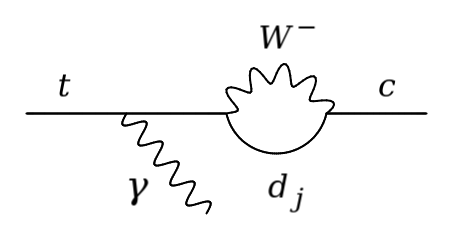}   
    \includegraphics[scale=0.25]{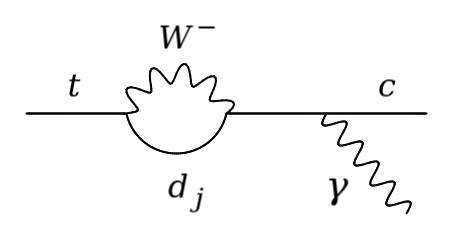} \\
    \includegraphics[scale=0.25]{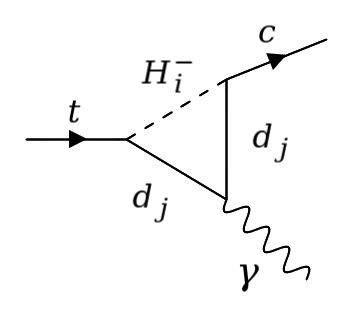}
    \includegraphics[scale=0.25]{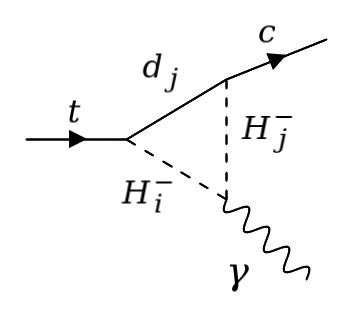}
    \includegraphics[scale=0.25]{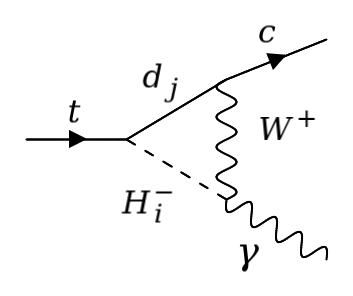}   
    \includegraphics[scale=0.25]{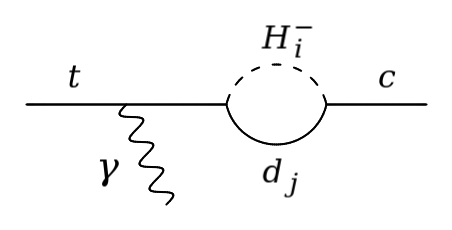}
    \includegraphics[scale=0.25]{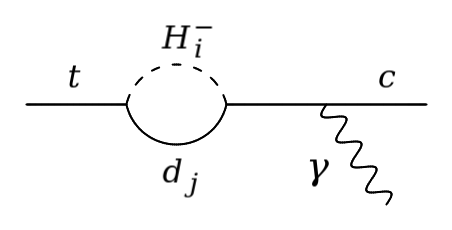} 
     \caption{Feynman diagrams contributing to the $t\to c\gamma$ process in the SM (top row) and in the 3HDM (bottom rows).\label{fig:tcgamma}}
\end{figure}

Multi-Higgs doublet models such as the democratic 3HDM considered here naturally provide additional scalar states that can mediate FCNC processes at loop level. In the alignment limit, one CP-even scalar possesses SM-like Higgs couplings, while the remaining non-SM scalars carry complementary interactions with fermions and gauge bosons. These non-SM scalars are thus the primary contributors to loop-induced top FCNC processes, as the SM-like Higgs is aligned and hence gives negligible contributions. Representative Feynman diagrams for the $t \rightarrow c \gamma$ transition in the SM and in the 3HDM are shown in Fig.~\ref{fig:tcgamma}, illustrating how charged scalars enter at the loop level in a 3HDM setting. To compute these loop-induced top FCNC amplitudes in the 3HDM, we proceed as follows. The model is implemented in \texttt{SARAH} \cite{Staub:2013tta, Goodsell:2014bna, Goodsell:2017pdq} to generate \texttt{FeynArts} \cite{Hahn:2000kx, Hahn:2010zi} compatible model files. This allows for the automatic generation of all relevant one-loop diagrams contributing to $t\to q X$ transitions, where $q = u,c$ and $X = \gamma, g, Z, H_{1,2,3}, A_{2,3}$. The corresponding amplitudes are then evaluated analytically using \texttt{FormCalc} \cite{Hahn:2010zi, Hahn:2016ebn} and numerically integrated with \texttt{LoopTools} \cite{Hahn:2010zi}. The corresponding rare top branching ratios are finally obtained by evaluating the ratio of the corresponding partial width to a top quark total width of $\Gamma_t=1.42$~GeV~\cite{ParticleDataGroup:2024cfk}. 

To quantify the possible enhancement of top FCNC branching ratios, we perform a numerical scan over a wide region of the 3HDM parameter space, covering the scalar masses, mixing angles and Yukawa sector parameters relevant for each alignment hierarchy. The scan is carried out independently for the three possible CP-even alignment configurations, in which each of the neutral CP-even scalars in turn plays the role of the SM-like Higgs boson. The multidimensional parameter space is sampled using Sobol quasi-random sequences~\cite{sobol1967distribution}, which ensures its efficient and uniform coverage while avoiding clustering effects typical of purely random scans. Throughout the scan, a set of theoretical consistency conditions is imposed to ensure the viability of the scalar sector. These include the vacuum stability of the scalar potential, the perturbativity of all scalar and Yukawa couplings and the tree-level unitarity of the multi-scalar scattering amplitudes. The vacuum stability requirement~\cite{Batra:2025amk} implies that the scalar potential is bounded from below in all directions of the field space, thereby preventing the emergence of unstable vacua. Meanwhile, perturbativity constraints are imposed on all quartic couplings appearing in Eqn.~\eqref{eq:scalarpot}, requiring $|\lambda_i|, |\lambda_{ij}|, |\lambda_{ij}'|, |\tilde\lambda_i| \leq 4\pi$ so that the theory remains perturbatively calculable. Following the results of Ref.~\cite{Akeroyd:2000wc, Boto:2021qgu, Bento:2022vsb}, we require that all 21 eigenvalues of the multi-scalar part of the scattering matrix are smaller than $8\pi$, thereby guaranteeing unitarity. 

Points satisfying these theoretical requirements are subsequently subjected to a comprehensive set of experimental constraints which are chosen to reflect the observables most sensitive to new scalar degrees of freedom. Among these, the electroweak precision tests encapsulated in the oblique parameters $S$, $T$ and $U$ are particularly sensitive to the mass splittings and mixing among the new scalar states. We thus compute these parameters using \texttt{SPheno}~\cite{Porod:2003um, Porod:2011nf} with the 3HDM model file implemented in \texttt{SARAH} so that all scalar contributions are included at one-loop, and we require the obtained values to lie within $1\sigma$ of the current experimental values~\cite{ParticleDataGroup:2024cfk},
\begin{equation}
        S = -0.02 \pm 0.10, \qquad
        T = 0.03 \pm 0.12\qquad\text{and} \qquad
        U = 0.01 \pm 0.11. 
\end{equation}
In addition, we impose direct exclusion limits on the additional Higgs states using results from searches at LEP, the Tevatron and the LHC, covering both additional neutral and charged scalars. We also ensure that the SM-like Higgs boson in the model reproduces the observed properties of the 125~GeV state probed at the LHC. This is achieved by computing Higgs signal strengths in various production and decay channels and performing a global goodness-of-fit test against the experimental measurements. All of these checks are implemented using the \texttt{HiggsTools} package~\cite{Bahl:2022igd}. Finally, constraints from flavour physics are included by imposing the most recent experimental bounds on the branching ratio $\mathcal{BR}(B \rightarrow X_s \gamma)$ that is particularly sensitive to the presence of additional charged scalars in the model's field content and the possible constructive and destructive interference between their contributions. Theoretical predictions for this decay are computed at next-to-leading order in QCD, following the methods depicted in Refs.~\cite{Akeroyd:2020nfj, Boto:2021qgu}, and the predicted branching ratio is required to lie within the $3\sigma$ experimental range~\cite{HeavyFlavorAveragingGroupHFLAV:2024ctg},
\begin{equation}
    2.87 \times 10^{-4} < \mathcal{BR}(B \rightarrow X_s \gamma) < 3.77 \times 10^{-4}.
\end{equation}

\subsection{Predictions for Top FCNC Branching Ratios in the Democratic 3HDM}
We now present predictions for FCNC top decays in the democratic 3HDM across the three possible CP-even scalar alignment scenarios. The results are obtained from two comprehensive parameter scans, which allow us to distinguish between the parameter space allowed solely by theoretical requirements and the more restricted region consistent with both theoretical and experimental constraints. The former reflects the intrinsic structure of the scalar sector, while the latter additionally encodes the impact of the assumed pattern for the Yukawa interactions. This distinction is important, as alternative Yukawa structures within the same $Z_3$-symmetric potential could modify the experimental constraints on the parameter space without affecting the underlying scalar spectrum. For each alignment scenario, we determine the range of branching ratios for the rare top decays $t \to q X$, and we analyse the impact of the extended scalar sector on these transitions. In particular, we highlight how the interplay between the scalar mass spectrum and the alignment scenario shapes the allowed regions of the parameter space and identify the decay channels most promising for experimental probes in future collider runs. 

\begin{table}\renewcommand{\arraystretch}{1.4}
    \centering
    \begin{tabular}{l l l}
    \toprule[1pt]
    Parameter & \ \ \ \ \ \ \ Scan Range (T)  & \ \ \ \ \ \ \ Scan Range (T+E)  \\
        \midrule[1pt]
        $m_{H_2^\pm}$ & \ \ \ \ \ \ \ $[200, 640]$ GeV & \ \ \ \ \ \ \ $[370, 570]$ GeV \\
        $m_{H_3^\pm}$ & \ \ \ \ \ \ \ $[200, 615]$ GeV & \ \ \ \ \ \ \ $[290, 440]$ GeV \\
        $\tan \beta_1$ & \ \ \ \ \ \ \ $[0.1, 5.6]$ & \ \ \ \ \ \ \ $[0.5, 2.5]$ \\
        $\tan \beta_2$ & \ \ \ \ \ \ \ $[0.1, 2.7]$  & \ \ \ \ \ \ \ $[0.5, 1.5]$ \\
        $\tan \theta$ & \ \ \ \ \ \ \ $[0.1, 100]$  & \ \ \ \ \ \ \ $[5.5, 100]$ \\
        $m_{H_2}$ & \ \ \ \ \ \ \ $[200, 640]$  GeV & \ \ \ \ \ \ \ $[360, 580]$ GeV \\
        $m_{H_3}$ & \ \ \ \ \ \ \ $[200, 615]$  GeV & \ \ \ \ \ \ \ $[200, 460]$ GeV \\
        $m_{A_2}$ & \ \ \ \ \ \ \ $[200, 650]$  GeV & \ \ \ \ \ \ \ $[350, 590]$ GeV \\
        $m_{A_3}$ & \ \ \ \ \ \ \ $[200, 630]$  GeV & \ \ \ \ \ \ \ $[200, 500]$ GeV \\
        $\tan \gamma$ & \ \ \ \ \ \ \ $[0.1, 100]$  & \ \ \ \ \ \ \ $[8, 100]$ \\
        $\tan \alpha_3$ & \ \ \ \ \ \ \ $[0.1, 100]$  & \ \ \ \ \ \ \ $[10, 100]$ \\
        \bottomrule[1pt]
    \end{tabular}
    \caption{Allowed parameter ranges for the regular hierarchy scenario. The label `T' corresponds to points allowed by theoretical constraints only, while `T+E' indicates points consistent with both theoretical and experimental bounds.\label{tab:al1-parameterscan-full}}
\end{table}

In the regular hierarchy scenario, the lightest CP-even scalar $H_1$ plays the role of the SM-like Higgs, while the heavier scalars $H_2$ and $H_3$ carry complementary interactions and contribute to flavour-changing processes at loop level. The parameter ranges resulting from our scans are collected in Table~\ref{tab:al1-parameterscan-full}, where the first column (`T') corresponds to points allowed by theoretical consistency only, while the second column (`T+E') additionally satisfies the experimental constraints discussed in Section~\ref{sec:framework}. The significant compression of the allowed mass ranges in the `T+E' scan relative to the theory-only case is primarily driven by the constraints originating from the oblique parameters. In multi-Higgs doublet models, they are highly sensitive to the mass splitting between the charged and neutral scalar states so that the electroweak precision constraints force the scan to converge into regions where the extra Higgs states are nearly mass-degenerate. This mass degeneracy, in turn, limits the size of the associated contributions to loop-induced FCNC amplitudes. Moreover, the requirement that $H_1$ has properties compatible with the LHC Higgs signal strengths imposes a near-alignment condition. We also notice that the `T+E' ranges for the scalar mixing parameters $\tan \gamma$ and $\tan \alpha_3$ are restricted to values greater than approximately 10, enforced by the direct BSM Higgs searches. In this regime, the non-SM scalars are thus sufficiently decoupled from the SM-like state, ensuring in particular that the production and decay rates of the neutral states in the $\gamma\gamma$, $ZZ$ and $WW$ channels remain consistent with the observations.

\begin{figure}
    \centering
    \includegraphics[width=0.75\columnwidth]{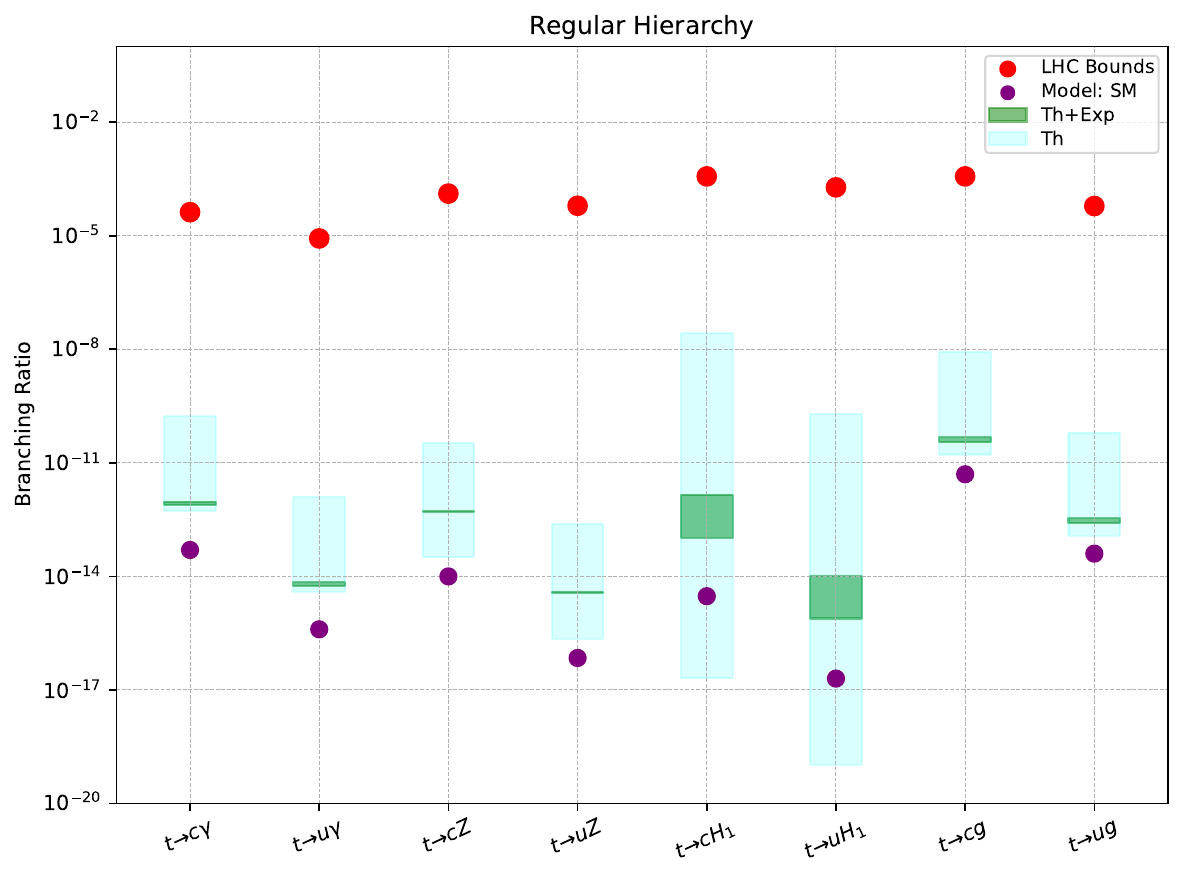}\\ \vspace*{.5cm}
    \renewcommand{\arraystretch}{1.4}
    \hspace*{0.8cm}\begin{tabular}{l l l l l}
    \toprule[1pt]
    Process & \ \ \ \ \ \ \ $\mathcal{BR}_{\max}$(T) & \ \ \ \ \ \ \ $\mathcal{BR}_{\min}$(T) & \ \ \ \ \ \ \ $\mathcal{BR}_{\max}$(T+E) & \ \ \ \ \ \ \ $\mathcal{BR}_{\min}$(T+E) \\
        \midrule[1pt]
        $t \rightarrow c\gamma$ & \ \ \ \ \ \ \ $1.73 \times 10^{-10}$ & \ \ \ \ \ \ \ $5.42 \times 10^{-13}$ & \ \ \ \ \ \ \ $9.51 \times 10^{-13}$ & \ \ \ \ \ \ \ $7.75 \times 10^{-13}$  \\
        $t \rightarrow cg$ & \ \ \ \ \ \ \ $8.66 \times 10^{-9}$ & \ \ \ \ \ \ \ $1.66 \times 10^{-11}$ & \ \ \ \ \ \ \ $4.73 \times 10^{-11}$ & \ \ \ \ \ \ \ $3.50 \times 10^{-11}$  \\
        $t \rightarrow cZ$ & \ \ \ \ \ \ \ $3.34 \times 10^{-11}$ & \ \ \ \ \ \ \ $3.21 \times 10^{-14}$ & \ \ \ \ \ \ \ $5.42 \times 10^{-13}$ & \ \ \ \ \ \ \ $5.05 \times 10^{-13}$  \\
        $t \rightarrow cH_1$ & \ \ \ \ \ \ \ $2.67 \times 10^{-8}$ & \ \ \ \ \ \ \ $2.04 \times 10^{-17}$ & \ \ \ \ \ \ \ $1.44 \times 10^{-12}$ & \ \ \ \ \ \ \ $1.06 \times 10^{-13}$  \\
        \midrule[1pt]
        $t \rightarrow u\gamma$ & \ \ \ \ \ \ \ $1.26 \times 10^{-12}$ & \ \ \ \ \ \ \ $3.94 \times 10^{-15}$ & \ \ \ \ \ \ \ $6.94 \times 10^{-15}$ & \ \ \ \ \ \ \ $5.65 \times 10^{-15}$  \\
        $t \rightarrow ug$ & \ \ \ \ \ \ \ $6.32 \times 10^{-11}$ & \ \ \ \ \ \ \ $1.20 \times 10^{-13}$ & \ \ \ \ \ \ \ $3.44 \times 10^{-13}$ & \ \ \ \ \ \ \ $2.54 \times 10^{-13}$  \\
        $t \rightarrow uZ$ & \ \ \ \ \ \ \ $2.43 \times 10^{-13}$ & \ \ \ \ \ \ \ $2.27 \times 10^{-16}$ & \ \ \ \ \ \ \ $3.94 \times 10^{-15}$ & \ \ \ \ \ \ \ $3.68 \times 10^{-15}$  \\
         $t \rightarrow uH_1$ & \ \ \ \ \ \ \ $1.94 \times 10^{-10}$ & \ \ \ \ \ \ \ $1.07 \times 10^{-19}$ & \ \ \ \ \ \ \ $1.04 \times 10^{-14}$ & \ \ \ \ \ \ \ $7.75 \times 10^{-16}$  \\
        \bottomrule[1pt]
    \end{tabular}
    \caption{Predicted branching ratios for top FCNC decays in the regular hierarchy scenario. Results are shown for scenarios consistent with theoretical requirements (`T', cyan) and additionally satisfying the imposed experimental constraints (`T+E', green). The upper panel represents the full scan, while the lower panel provides numerical values for the maxima and minima for each decay channel. \label{fig:reghierarchy}}
\end{figure}

The upper panel of Figure~\ref{fig:reghierarchy} presents the predicted branching ratios, with the cyan band corresponding to points allowed by theoretical constraints only, while the green band satisfies both theoretical and current experimental constraints. The red markers indicate the current LHC upper limits, and the violet markers denote the SM predictions. For completeness, the lower panel of the figure reports the numerical maxima and minima of the branching ratios for each decay channel, as obtained from the scans. Predicted branching ratios for top FCNC decays span several orders of magnitude, reflecting the interplay of the scalar masses, mixing angles and the Yukawa/gauge couplings within the $Z_3$-symmetric 3HDM. Across the entire scanned parameter space, we observe a consistent hierarchy in branching ratios: $\mathcal{BR}(t \to q g) > \mathcal{BR}(t \to q \gamma) > \mathcal{BR}(t \to q Z)$, as expected from the relative strengths of the gauge couplings involved. Crucially, the 3HDM loop contributions bypass the GIM-suppression present in the SM, resulting in 3HDM rates that are several orders of magnitude higher than in the SM. In the theory-only scan, the dominant channels $t \to q \gamma$ and $t \to q g$ typically lie between $10^{-13}$ and $10^{-9}$, while the inclusion of experimental constraints reduces these ranges by roughly two to four orders of magnitude. The $t \to q H_1$ mode exhibits a wider spread, occasionally dipping below the SM expectation in theory-favourable regions of the parameter space while remaining enhanced relative to the SM in most of the parameter space. This wide variation ($10^{-8}$ down to $10^{-19}$) directly probes the degree of alignment, with the lower edge corresponding to configurations effectively indistinguishable from the SM. Finally, these results indicate that although present data do not yet probe this region of the 3HDM parameter space, the upper edge of the predicted bands could become accessible at future collider experiments.

\begin{table}\renewcommand{\arraystretch}{1.4}
    \centering
    \begin{tabular}{l l l}
    \toprule[1pt]
    Parameter & \ \ \ \ \ \ \ Scan Range (T) & \ \ \ \ \ \ \ Scan Range (T+E) \\
        \midrule[1pt]
        $m_{H_2^\pm}$ & \ \ \ \ \ \ \ $[100, 620]$ GeV & \ \ \ \ \ \ \ $[345, 440]$ GeV \\
        $m_{H_3^\pm}$ & \ \ \ \ \ \ \ $[100, 620]$ GeV & \ \ \ \ \ \ \ $[230, 520]$ GeV \\
        $\tan \beta_1$ & \ \ \ \ \ \ \ $[0.1, 15.5]$  & \ \ \ \ \ \ \ $[0.5, 5]$ \\
        $\tan \beta_2$ & \ \ \ \ \ \ \ $[0.1, 6]$  & \ \ \ \ \ \ \ $[1, 3.5]$ \\
        $\tan \theta$ & \ \ \ \ \ \ \ $[0.1, 100]$  & \ \ \ \ \ \ \ $[0.6, 100]$ \\
        $m_{H_1}$ & \ \ \ \ \ \ \ $[60, 120]$ GeV  & \ \ \ \ \ \ \ $[80, 120]$ GeV \\
        $m_{H_3}$ & \ \ \ \ \ \ \ $[200,  620]$ GeV  & \ \ \ \ \ \ \ $[370, 590]$ GeV \\
        $m_{A_2}$ & \ \ \ \ \ \ \ $[80,  630]$ GeV  & \ \ \ \ \ \ \ $[197, 340]$ GeV \\
        $m_{A_3}$ & \ \ \ \ \ \ \ $[80,  630]$ GeV  & \ \ \ \ \ \ \  $[80, 210]$ GeV \\
        $\tan \gamma$ & \ \ \ \ \ \ \ $[0.1, 100]$ & \ \ \ \ \ \ \  $[3.4, 100]$ \\
    \bottomrule[1pt]
    \end{tabular}
    \caption{Allowed parameter ranges for the medial hierarchy scenario. The label `T' corresponds to points allowed by theoretical constraints only, while `T+E' indicates points consistent with both theoretical and experimental bounds. \label{tab:al2-parameterscan-full}}
\end{table}

In the medial hierarchy scenario, the intermediate CP-even scalar $H_2$ takes the role of the SM-like Higgs, while the lighter $H_1$ and heavier $H_3$ states carry complementary interactions. The parameter ranges resulting from our scans of this 3HDM setup are collected in Table~\ref{tab:al2-parameterscan-full}, where once again the first column `T' corresponds to points allowed solely by theoretical consistency and the second column `T+E' additionally includes the imposed experimental constraints. As in the regular hierarchy case, the oblique parameters are highly sensitive to the mass splittings among the charged and neutral scalars, and the `T+E' scan converges to regions with approximate mass degeneracy. Additionally, the requirement that $H_2$ reproduces the observed LHC Higgs properties enforces a near-alignment condition for this state, which is reflected in the more restricted `T+E' range of $\tan \gamma\gtrsim 3.4$. By orthogonality of the scalar mixing matrix, the remaining CP-even states $H_1$ and $H_3$ reside largely in the subspace orthogonal to the vev direction. In the Higgs basis, this implies suppressed couplings to electroweak gauge bosons but unsuppressed Yukawa interactions. Equivalently, the $t \to q H_1$ branching ratio is not expected to be constrained by the alignment conditions imposed on the SM-like 125~GeV Higgs $H_2$.

\begin{figure}
    \centering
    \includegraphics[width=0.75\columnwidth]{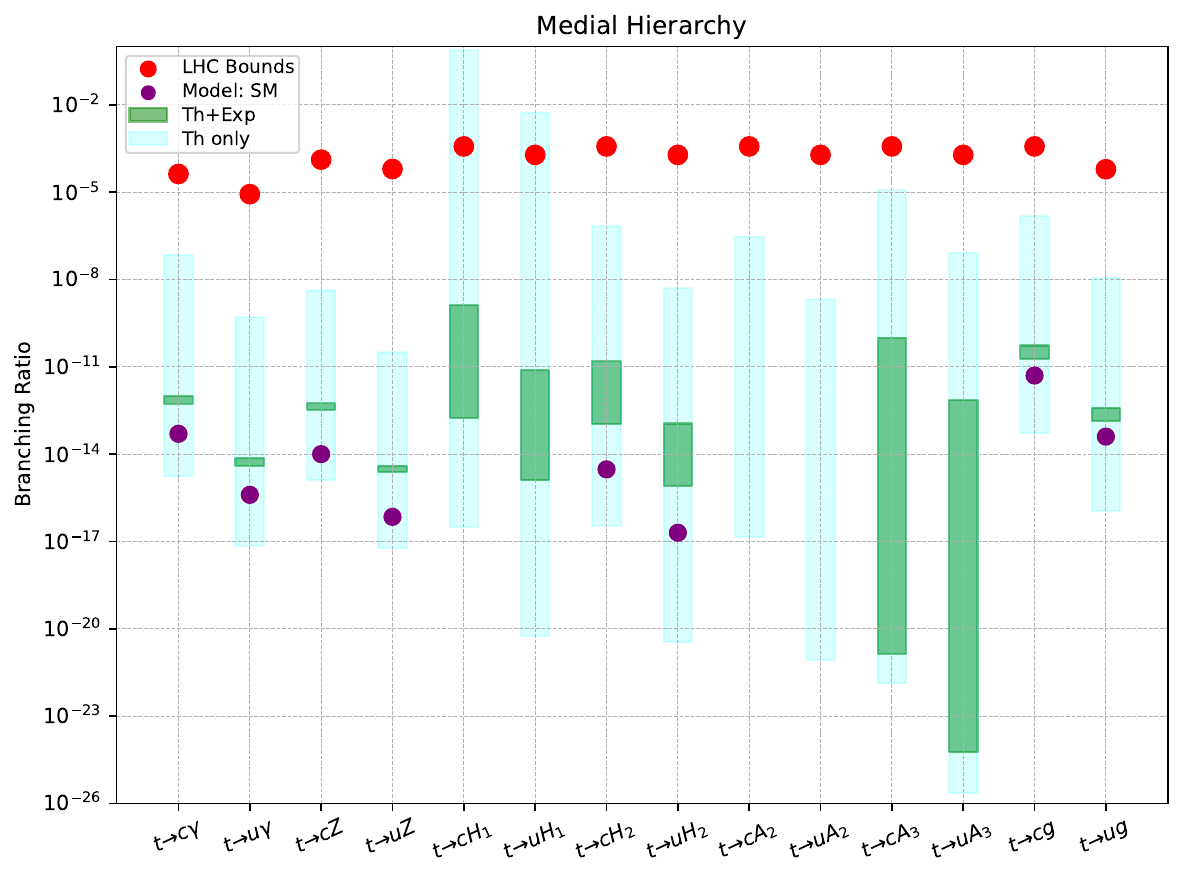}\vspace*{.5cm}
    \renewcommand{\arraystretch}{1.4}
    \hspace*{0.8cm}\begin{tabular}{l l l l l}
    \toprule[1pt]
    Process & \ \ \ \ \ \ \ $\mathcal{BR}_{max}$(T) & \ \ \ \ \ \ \ $\mathcal{BR}_{min}$(T) & \ \ \ \ \ \ \ $\mathcal{BR}_{max}$(T+E) & \ \ \ \ \ \ \ $\mathcal{BR}_{min}$(T+E) \\
        \midrule[1pt]
        $t \rightarrow c\gamma$ & \ \ \ \ \ \ \ $6.93 \times 10^{-8}$ & \ \ \ \ \ \ \ $1.73 \times 10^{-15}$ & \ \ \ \ \ \ \ $1.01 \times 10^{-12}$ & \ \ \ \ \ \ \ $5.42 \times 10^{-13}$  \\
        $t \rightarrow cg$ & \ \ \ \ \ \ \ $1.52 \times 10^{-6}$ & \ \ \ \ \ \ \ $5.25 \times 10^{-14}$ & \ \ \ \ \ \ \ $5.40 \times 10^{-11}$ & \ \ \ \ \ \ \ $1.87 \times 10^{-11}$  \\
        $t \rightarrow cZ$ & \ \ \ \ \ \ \ $4.16 \times 10^{-9}$ & \ \ \ \ \ \ \ $1.25 \times 10^{-15}$ & \ \ \ \ \ \ \ $5.51 \times 10^{-13}$ & \ \ \ \ \ \ \ $3.28 \times 10^{-13}$  \\
        $t \rightarrow cH_1$ & \ \ \ \ \ \ \ $7.32 \times 10^{-1}$ & \ \ \ \ \ \ \ $3.13 \times 10^{-17}$ & \ \ \ \ \ \ \ $1.30 \times 10^{-9}$ & \ \ \ \ \ \ \ $1.73 \times 10^{-13}$  \\
        $t \rightarrow cH_2$ & \ \ \ \ \ \ \ $6.96 \times 10^{-7}$ & \ \ \ \ \ \ \ $3.50 \times 10^{-17}$ & \ \ \ \ \ \ \ $1.54 \times 10^{-11}$ & \ \ \ \ \ \ \ $1.08 \times 10^{-13}$\\
        $t \rightarrow cA_2$ & \ \ \ \ \ \ \ $2.92 \times 10^{-7}$  & \ \ \ \ \ \ \ $1.43 \times 10^{-17}$ & \ \ \ \ \ \ \ $-$  & \ \ \ \ \ \ \ $-$ \\
        $t \rightarrow cA_3$ & \ \ \ \ \ \ \ $1.14 \times 10^{-5}$ & \ \ \ \ \ \ \ $1.35 \times 10^{-22}$ & \ \ \ \ \ \ \ $1.0 \times 10^{-10}$ & \ \ \ \ \ \ \ $1.38\times10^{-21}$ \\
        \midrule[1pt]
        $t \rightarrow u\gamma$ & \ \ \ \ \ \ \ $5.04 \times 10^{-10}$ & \ \ \ \ \ \ \ $7.25 \times 10^{-18}$ & \ \ \ \ \ \ \ $7.39 \times 10^{-15}$ & \ \ \ \ \ \ \ $3.94 \times 10^{-15}$  \\
        $t \rightarrow ug$ & \ \ \ \ \ \ \ $1.14 \times 10^{-8}$ & \ \ \ \ \ \ \ $1.12 \times 10^{-16}$ & \ \ \ \ \ \ \ $3.93 \times 10^{-13}$ & \ \ \ \ \ \ \ $1.36 \times 10^{-13}$  \\
        $t \rightarrow uZ$ & \ \ \ \ \ \ \ $3.10 \times 10^{-11}$ & \ \ \ \ \ \ \ $5.93 \times 10^{-18}$ & \ \ \ \ \ \ \ $4.01 \times 10^{-15}$ & \ \ \ \ \ \ \ $2.39 \times 10^{-15}$  \\
         $t \rightarrow uH_1$ & \ \ \ \ \ \ \ $5.35 \times 10^{-3}$ & \ \ \ \ \ \ \ $5.82 \times 10^{-21}$ & \ \ \ \ \ \ \ $7.89 \times 10^{-12}$ & \ \ \ \ \ \ \ $1.26 \times 10^{-15}$  \\
        $t \rightarrow uH_2$ & \ \ \ \ \ \ \ $5.02 \times 10^{-9}$ & \ \ \ \ \ \ \ $3.42 \times 10^{-21}$ & \ \ \ \ \ \ \ $1.13 \times 10^{-13}$ & \ \ \ \ \ \ \ $7.89 \times 10^{-16}$\\
        $t \rightarrow uA_2$ & \ \ \ \ \ \ \ $2.12 \times 10^{-9}$  & \ \ \ \ \ \ \ $8.45 \times 10^{-22}$ & \ \ \ \ \ \ \ $-$  & \ \ \ \ \ \ \ $-$ \\
        $t \rightarrow uA_3$ & \ \ \ \ \ \ \ $8.31 \times 10^{-8}$ & \ \ \ \ \ \ \ $2.22 \times 10^{-26}$ & \ \ \ \ \ \ \ $7.18 \times 10^{-13}$ & \ \ \ \ \ \ \ $5.73\times10^{-25}$ \\
    \bottomrule[1pt]
    \end{tabular}
    \caption{Predicted branching ratios for top FCNC decays in the medial hierarchy scenario. Results are shown for scenarios consistent with theoretical requirements (`T', cyan) and additionally satisfying the imposed experimental constraints (`T+E', green). The upper panel represents the full scan, while the lower panel provides numerical values for the maxima and minima for each decay channel.\label{fig:med-hierarchy}}
\end{figure}

The predicted top FCNC branching ratios are shown in Fig.~\ref{fig:med-hierarchy}, with the colour coding following the same convention as for the regular hierarchy case: the cyan band corresponds to scan points consistent with the imposed theoretical requirements, while the green band includes points satisfying both theoretical and experimental constraints. In addition, we remind that the experimental limits on the rare decays $t \to q H_2$ and $t \to q A_{2,3}$ can be derived under the assumption that the final-state Higgs decays to $b\bar{b}$. We can thus recast these limits for the non-SM states by rescaling the bounds according to the ratios of the relevant 3HDM Yukawa couplings to the corresponding SM values. This procedure then ensures that the predicted branching ratios for the $t \to q H_1$ and $t \to q A_{2,3}$ decays can be compared with current LHC data. Finally, the lower panel of the figure reports the numerical maxima and minima of the branching ratios for each channel as extracted from the scans. 

Across the scanned parameter space, the relative size of the FCNC branching ratios remains $\mathcal{BR}(t \to q g) > \mathcal{BR}(t \to q \gamma) > \mathcal{BR}(t \to q Z)$, as in the regular hierarchy. However, the lighter non-SM scalar $H_1$ and the additional light CP-odd states $A_2$ and $A_3$ open new decay channels. In the viable regions of the parameter space, the $t \to q H_1$ mode can reach branching ratios as high as $\mathcal{O}(10^{-1})$, reflecting the sizeable couplings of the light scalar to the top quark and the absence of alignment suppression. This contrasts with the $t \to q H_2$ decay that is constrained by alignment and LHC signal-strength data. On the other hand, the $t \to q A_2$ decay becomes kinematically forbidden in the `T+E' scan, which explains the absence of a green band for this channel in Fig.~\ref{fig:med-hierarchy}. This is a direct consequence of the correlated scalar mass spectrum enforced by the $Z_3$-symmetric potential. Requiring a SM-like CP-even scalar $H_2$ at 125 GeV together with electroweak precision constraints favours spectra in which $A_2$ is relatively heavy. On the contrary, the remaining pseudoscalar $A_3$ belongs to the non-SM-like sector and can thus remain significantly lighter without violating any electroweak bounds, thereby allowing the decay channel $t \to q A_3$ to remain kinematically accessible in the `T+E' scan but with suppressed branching ratios.

The inclusion of experimental constraints when exploring medial hierarchy scenarios in the 3HDM parameter space substantially reduces the allowed branching ratios for all channels, typically by several orders of magnitude. For example, in the `T+E' scan, the dominant decays $t \to q \gamma$ and $t \to q g$ are restricted to $\mathcal{O}(10^{-13}-10^{-11})$, while the $t \to q H_1$ and $t \to q H_2$ modes are pushed down to $\mathcal{O}(10^{-13}-10^{-9})$, highlighting the strong impact of alignment and Higgs mass requirements. The spread in $\mathcal{BR}(t \to q H_2)$ again serves as a numerical probe of the degree of alignment, in close analogy with the behaviour observed in the regular hierarchy scenario: the smaller values allowed for the associated branching ratio correspond to configurations effectively indistinguishable from the SM, while the largest ones reflect regions of the parameter space  with substantial deviations from the SM in the scalar couplings. Notably, the lighter non-SM (pseudo)scalars in the medial hierarchy enhance the sensitivity of some FCNC channels to scalar masses and mixing angles, offering potentially observable signatures in future collider experiments such as the HL-LHC.

\begin{table}
    \renewcommand{\arraystretch}{1.4}
    \centering
    \begin{tabular}{l l}
    \toprule[1pt]
    Parameter & \ \ \ \ \ \ \ Scan Range (T) \\
        \midrule[1pt]
        $m_{H_2^\pm}$ & \ \ \ \ \ \ \ $[80, 615]$  \\
        $m_{H_3^\pm}$ & \ \ \ \ \ \ \ $[80, 611]$  \\
        $\tan \beta_1$ & \ \ \ \ \ \ \ $[0.5, 5]$  \\
        $\tan \beta_2$ & \ \ \ \ \ \ \ $[1, 7.8]$  \\
        $\tan \theta$ & \ \ \ \ \ \ \ $[0.1, 100]$  \\
        $m_{H_1}$ & \ \ \ \ \ \ \ $[60, 120]$  \\
        $m_{H_2}$ & \ \ \ \ \ \ \ $[60, 120]$  \\
        $m_{A_2}$ & \ \ \ \ \ \ \ $[60, 362]$ \\
        $m_{A_3}$ & \ \ \ \ \ \ \ $[60, 570]$  \\
        $\tan \gamma$ & \ \ \ \ \ \ \ $[0.5, 100]$ \\
    \bottomrule[1pt]
    \end{tabular} 
    \caption{Allowed parameter ranges for the inverted hierarchy scenario for points allowed by theoretical constraints only. \label{tab:al3-parameterscan-full}}
\end{table}

\begin{figure}
    \centering
    \includegraphics[width=0.75\columnwidth]{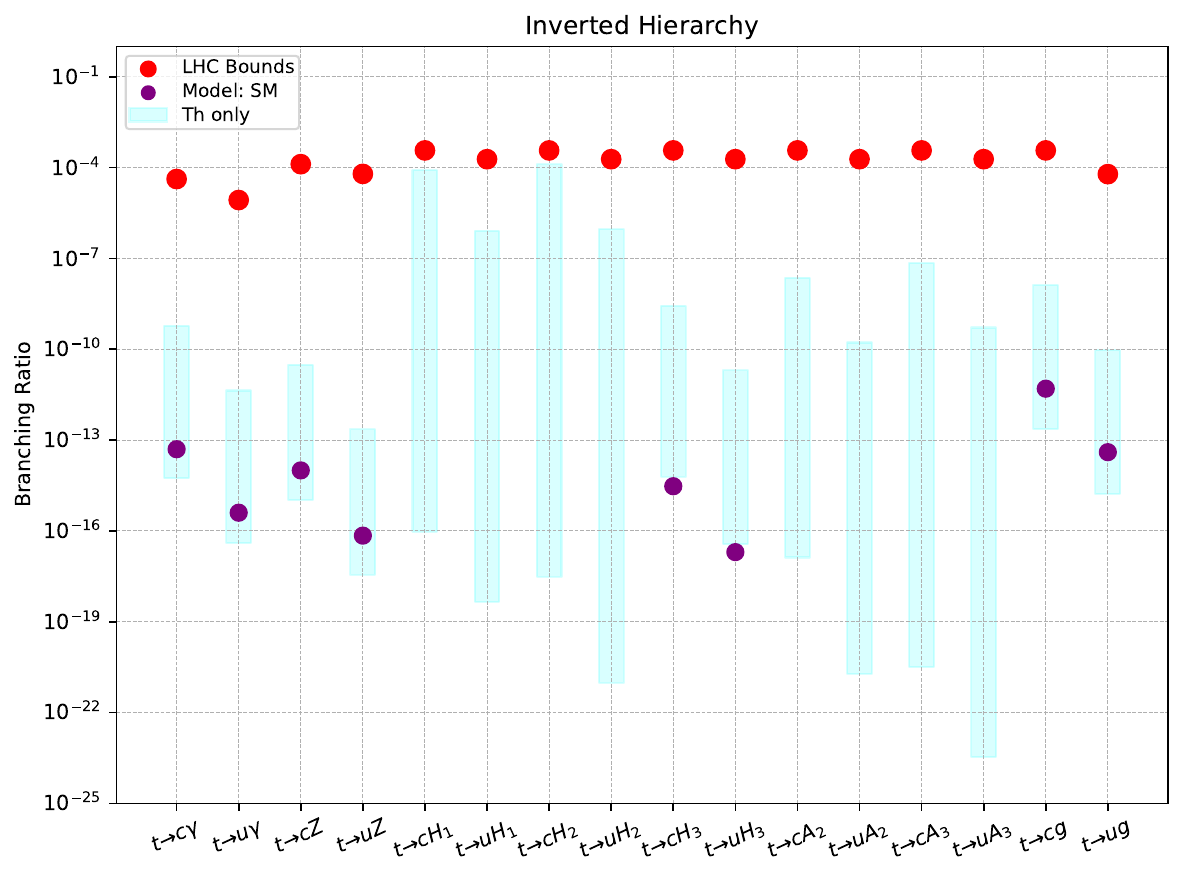}\vspace*{.5cm}
    \renewcommand{\arraystretch}{1.4}
    \begin{tabular}{l l l}
    \toprule[1pt]
    Process & \ \ \ \ \ \ \ $\mathcal{BR}_{max}$ & \ \ \ \ \ \ \ $\mathcal{BR}_{min}$ \\
        \midrule[1pt]
        $t \rightarrow c\gamma$ & \ \ \ \ \ \ \ $5.99 \times 10^{-10}$ & \ \ \ \ \ \ \ $5.44 \times 10^{-15}$  \\
        $t \rightarrow cg$ & \ \ \ \ \ \ \ $1.34 \times 10^{-8}$ & \ \ \ \ \ \ \ $2.31 \times 10^{-13}$  \\
        $t \rightarrow cZ$ & \ \ \ \ \ \ \ $3.13 \times 10^{-11}$ & \ \ \ \ \ \ \ $1.04 \times 10^{-15}$  \\
        $t \rightarrow cH_1$ & \ \ \ \ \ \ \ $8.38 \times 10^{-5}$ & \ \ \ \ \ \ \ $9.58 \times 10^{-17}$  \\
        $t \rightarrow cH_2$ & \ \ \ \ \ \ \ $1.28 \times 10^{-4}$ & \ \ \ \ \ \ \ $3.12 \times 10^{-18}$\\
        $t \rightarrow cH_3$ & \ \ \ \ \ \ \ $2.75 \times 10^{-9}$ & \ \ \ \ \ \ \ $6.26 \times 10^{-15}$\\
        $t \rightarrow cA_2$ & \ \ \ \ \ \ \ $2.32 \times 10^{-8}$  & \ \ \ \ \ \ \ $1.33\times10^{-17}$ \\
        $t \rightarrow cA_3$ & \ \ \ \ \ \ \ $7.18 \times 10^{-8}$ & \ \ \ \ \ \ \ $3.18\times10^{-21}$ \\
        \midrule[1pt]
        $t \rightarrow u\gamma$ & \ \ \ \ \ \ \ $4.36 \times 10^{-12}$ & \ \ \ \ \ \ \ $3.96 \times 10^{-17}$  \\
        $t \rightarrow ug$ & \ \ \ \ \ \ \ $9.72 \times 10^{-11}$ & \ \ \ \ \ \ \ $1.68 \times 10^{-15}$  \\
        $t \rightarrow uZ$ & \ \ \ \ \ \ \ $2.27 \times 10^{-13}$ & \ \ \ \ \ \ \ $3.60 \times 10^{-18}$  \\
         $t \rightarrow uH_1$ & \ \ \ \ \ \ \ $8.24 \times 10^{-7}$ & \ \ \ \ \ \ \ $4.58 \times 10^{-19}$  \\
        $t \rightarrow uH_2$ & \ \ \ \ \ \ \ $9.37 \times 10^{-7}$ & \ \ \ \ \ \ \ $9.72 \times 10^{-22}$\\
        $t \rightarrow uH_3$ & \ \ \ \ \ \ \ $2.01 \times 10^{-11}$ & \ \ \ \ \ \ \ $3.62 \times 10^{-17}$\\
        $t \rightarrow uA_2$ & \ \ \ \ \ \ \ $1.68 \times 10^{-10}$  & \ \ \ \ \ \ \ $1.90\times10^{-21}$ \\
        $t \rightarrow uA_3$ & \ \ \ \ \ \ \ $5.23 \times 10^{-10}$ & \ \ \ \ \ \ \ $3.38\times10^{-24}$ \\
    \bottomrule[1pt]
    \end{tabular}
    \caption{Predicted branching ratios for top FCNC decays in the inverted hierarchy scenario. Results are shown for scenarios consistent with theoretical requirements (`T', cyan) with the the upper panel representing the full scan and the lower panel providing the associated numerical values for the maxima and minima for each decay channel. \label{fig:inv-hierarchy}}
\end{figure}

In the inverted hierarchy scenario, the heaviest CP-even scalar $H_3$ plays the role of the SM-like Higgs boson, while the lighter states $H_1$ and $H_2$ populate the non-SM-like sector. The parameter ranges spanned by the viable scenarios in this configuration are collected in Table~\ref{tab:al3-parameterscan-full}, where we report the scan intervals for points satisfying theoretical consistency requirements only. In contrast to the regular and medial hierarchies, we find that no region of the inverted hierarchy parameter space simultaneously satisfies both theoretical constraints and current experimental bounds. As a result, only the theoretically allowed scan (`T') is shown in Fig.~\ref{fig:inv-hierarchy} through the cyan band. The absence of a viable `T+E' region is a direct consequence of the tension between the inverted mass ordering and the requirement of a SM-like Higgs boson at 125~GeV. In this scenario, the SM-like state $H_3$ is the heaviest CP-even scalar, which generically induces sizeable mass splittings between $H_3$, the lighter CP-even scalars $H_{1,2}$ and the charged and CP-odd states. Such splittings are however strongly constrained by electroweak precision tests and cannot be efficiently mitigated by alignment alone~\cite{Batra:2025amk}. In addition, flavour constraints further disfavour the inverted hierarchy configuration, as the presence of at least one relatively light charged Higgs boson can lead to sizeable contributions to the $B \to X_s \gamma$ decay, unless accidental cancellations among different scalar contributions occur. 

Despite the lack of phenomenologically viable points after imposing the considered experimental constraints, the inverted hierarchy remains instructive from a theoretical perspective. In the allowed `T’ scan region, all neutral scalars lie below the top-quark mass, thus rendering all FCNC decays $t \to q H_{1,2,3}$ and $t \to q A_{2,3}$ kinematically accessible. Consequently, all neutral scalars contribute to flavour-changing top decays, leading to a rich FCNC pattern. This is clearly reflected in Fig.~\ref{fig:inv-hierarchy}, where sizeable branching ratios are obtained for the $t \to q H_{1,2}$ and $t \to q A_{2,3}$ modes, in particular with $\mathcal{BR}(t \to c H_{1,2})$ approaching $\mathcal{O}(10^{-4})$ and $\mathcal{BR}(t \to c A_{2,3})$ reaching up to $\mathcal{O}(10^{-8})$. These enhanced scalar-mediated FCNC rates originate from two combined effects. First, the absence of a strict alignment requirement on the lighter CP-even scalars allows for sizeable flavour-violating Yukawa couplings, and second, the kinematic accessibility of all neutral scalars enhances the available phase space. The SM-like Higgs $H_3$, by contrast, exhibits more suppressed FCNC branching ratios, which is consistent with its role as the aligned state and the corresponding constraints on its Yukawa structure. Although the inverted hierarchy possibility is excluded once experimental constraints are imposed, the results shown here highlight the crucial role played by the scalar mass ordering in shaping the FCNC top-quark phenomenology. In particular, this scenario illustrates how large FCNC branching ratios can naturally arise in extended Higgs sectors when multiple light neutral scalars are present, even if such configurations are ultimately ruled out by precision data. As such, the inverted hierarchy therefore provides a useful theoretical benchmark against which the phenomenological viability of the regular and medial hierarchies can be assessed.

\section{Summary and outlook }
\label{sec:conclusion}

In this work, we have presented a comprehensive study of FCNC top-quark decays in the democratic 3HDM with a $Z_3$-symmetric scalar potential and Natural Flavour Conservation in the Yukawa sector. The presence of an extended Higgs spectrum comprising multiple CP-even, CP-odd and charged scalar states generically induces new one-loop contributions to rare top decays, rendering these processes a sensitive probe of both the scalar mass spectrum and the alignment structure of the model. We have performed extensive scans of the associated multidimensional parameter space, imposing theoretical consistency conditions as well as current collider and flavour constraints. Moreover, our analysis systematically explored the three possible CP-even scalar alignment configurations, the so-called regular, medial and inverted hierarchies, and evaluated the corresponding branching ratios for the FCNC processes $t \to q X$ with $q=u,c$ and $X=\gamma, g, Z, H_{1,2,3}, A_{2,3}$. Across all scenarios, a clear and robust pattern emerges: the SM-like Higgs state remains aligned with the electroweak vacuum and consequently exhibits strongly suppressed flavour-changing couplings, while the non-SM scalars carry the complementary interaction strength that governs the size of the FCNC effects.

In the regular hierarchy setting, where the lightest CP-even scalar is SM-like and the remaining scalars are typically heavier than the top quark, the FCNC branching ratios remain relatively suppressed. In this case, all rare top decays are found to lie safely below current experimental bounds, although several channels could approach the sensitivity expected at future facilities.  Scenarios with a medial hierarchy exhibit a qualitatively different behaviour. Here, the SM-like Higgs is the intermediate CP-even state, while lighter non-SM CP-even and CP-odd scalars can be present without violating current constraints. As a result, top decays into non-standard Higgs bosons can reach significantly enhanced branching ratios compared to scenarios with a regular hierarchy. In particular, decays of the form $t \to q H_1$ can approach values close to or even larger than current experimental sensitivities, highlighting the phenomenological importance of searches targeting non-SM Higgs final states in top-quark decays as potential discovery channels of the model. This alignment pattern hence illustrates how the alignment of the observed 125 GeV Higgs does not preclude sizeable flavour-violating effects mediated by additional states. Finally, setups with an inverted hierarchy in which the SM-like Higgs is the heaviest CP-even scalar do not admit parameter regions simultaneously consistent with all theoretical and experimental constraints. Nevertheless, the theoretically allowed region provides a useful benchmark for understanding the impact of the scalar mass ordering and mixings  on the FCNC phenomenology. In this configuration, all neutral scalars are indeed kinematically accessible in top decays, leading to potentially large FCNC branching ratios driven by the absence of strong alignment suppression for the lighter states.

Taken together, our results demonstrate that FCNC top decays constitute a powerful and incisive probe of models with an extended Higgs sector. While the predicted branching ratios remain compatible with current experimental limits, several channels, particularly those involving decays into non-SM scalar states, lead to branching ratios tantalisingly close to the projected sensitivity of future collider programs. The High-Luminosity LHC, and even more so the next-generation hadron colliders, will therefore be capable of probing a substantial part of the phenomenologically viable regions of the parameter space identified in this work. Moreover, continued experimental scrutiny of rare top-quark decays will provide valuable insight into the structure of the scalar sector and the possible role of an extended Higgs dynamics in electroweak symmetry breaking.


\begin{acknowledgments}
AK gratefully acknowledges M.S.A.~Alam~Khan for insightful discussions and valuable inputs during the course of this work, and acknowledges the support from a Director's Fellowship at IIT Gandhinagar. The work of BF was supported in part by Grant ANR-21-CE31-0013 (Project DMwithLLPatLHC) from the French \textit{Agence Nationale de la Recherche}.
\end{acknowledgments}

\bibliographystyle{JHEP}
\bibliography{bibliography}

\end{document}